\documentclass{cpbtex}
\usepackage{booktabs}
\usepackage{graphicx}
\usepackage{amssymb}
\usepackage{lineno}
\usepackage{cmap}

\begin{document}

\title{Physical parameter regression from black hole images via a multiscale adaptive neural network}

\author{
  Jialei Wei$^{1}$, 
  Ao Liu$^{1}$\thanks{Corresponding author. E-mail: liuao@hunnu.edu.cn}, 
  Dejiang Li$^{1}$, 
  and Cuihong Wen$^{1}$\thanks{Corresponding author. E-mail: cuihongwen@hunnu.edu.cn\\
  This work was supported by the Natural Science Foundation of Hunan Province under grant No. 2023JJ30384; the National Natural Science Foundation of China under Grants No. 12374408; the University Students' Innovation and Entrepreneurship Training Program Project.} \\
  $^{1}$College of Information Science and Engineering, Hunan Normal University, \\ Changsha 410081, China
}

\maketitle

\begin{abstract}

High-precision regression of physical parameters from black hole images generated by General Relativistic Ray Tracing (GRRT) is essential for investigating spacetime curvature and advancing black hole astrophysics. However, due to limitations in observational resolution, high observational costs, and imbalanced distributions of positive and negative samples, black hole images often suffer from data scarcity, sparse parameter spaces, and complex structural characteristics. These factors pose significant challenges to conventional regression methods based on simplified physical models. To overcome these challenges, this study introduces Multiscale Adaptive Network (MANet) , a novel regression framework grounded in deep learning. MANet integrates an Adaptive Channel Attention (ACA) module to selectively enhance features in physically informative regions. Meanwhile, a Multiscale Enhancement Feature Pyramid (MEFP) is employed to capture fine-grained spatial structures such as photon rings and accretion disks, while alleviating information loss due to downsampling. Experimental evaluations on GRRT-simulated datasets demonstrate that MANet substantially improves parameter estimation accuracy and generalization capability in high-dimensional parameter spaces, outperforming existing baseline approaches. This framework presents a promising avenue for high-precision parameter regression in Event Horizon Telescope (EHT) data analysis and broader astrophysical imaging applications characterized by sparse and noisy data.

\end{abstract}



\section{Introduction}

In recent years, the Event Horizon Telescope (EHT) successfully imaged the supermassive black holes Sagittarius A* and M87\cite{M87,A*,A*2}, providing valuable observational evidence for the study of strong gravitational fields and general relativity. These images reveal the black hole shadow, photon rings, and accretion disk radiation structures, which contain information about spacetime geometry, plasma conditions, and gravitational dynamics\cite{Dexter2010,yy1,yy2,yy3,Narayan2023}. Inverting physical parameters from such images—such as black hole spin, accretion disk thickness, and electron temperature—not only helps reconstruct electromagnetic processes near the black hole but also offers important tests of general relativity and insights into black hole formation and evolution\cite{Johannsen2016,Chen2023,Younsi2023}.

With the development of General Relativistic Ray Tracing (GRRT) simulations\cite{GRRT,GRRT2,GRRT_review}, researchers can generate high-fidelity synthetic black hole images for modeling purposes. However, the limited observational resolution and high cost of data acquisition have resulted in sparse datasets and a highly nonlinear parameter space, both of which pose significant challenges to accurate regression. Traditional regression methods predominantly rely on physics-based models that compare GRRT-simulated images with observational data to infer black hole parameters. Common approaches include forward modeling combined with grid search, where parameter spaces are exhaustively sampled to identify best-fit models\cite{ref1}; and Bayesian inference methods, which incorporate prior knowledge and probabilistic frameworks to estimate posterior distributions of parameters\cite{Bayes,Bayes2}. Although these methods provide interpretability, they depend on idealized assumptions about accretion disk structure, magnetic field configuration, and radiation mechanisms, which may not fully capture the complexities of real observations\cite{ref3,ref4,You2023,Ebi2006}. Moreover, the computational cost of dense sampling in high-dimensional parameter spaces limits the efficiency and scalability of these methods. Under conditions of limited resolution and noisy data, such approaches are sensitive to initialization and may converge to local optima or produce nonphysical results. The nonlinear coupling of black hole image features further complicates explicit modeling, restricting the ability to capture detailed structures and maintain regression robustness\cite{Guo2024}.

Deep learning approaches have demonstrated strong capabilities in nonlinear modeling and feature extraction across various computer vision tasks, such as classification, recognition, and regression\cite{DL_review,Khan2022,Wei2022,Save2022,Le2022}. Convolutional neural networks (CNNs) can learn complex mappings between images and parameters from data, improving prediction accuracy and efficiency\cite{AlexNet}. Compared with traditional physics-based methods, deep learning reduces modeling bias and improves performance in the presence of observational noise and limited data, making it particularly suitable for black hole image-based parameter estimation\cite{DL_book,EHT_L4,Patel2022}.

However, the particular characteristics of black hole images pose challenges for training and generalization. Key physical structures such as photon rings\cite{Johnson2020} are subtle and sensitive to noise, complicating feature extraction\cite{Psaltis2020,jxjg}. Additionally, current GRRT-based synthetic datasets are relatively small for deep network training, increasing the risk of overfitting and limiting the ability to capture complex physical phenomena. Therefore, it is essential to design an efficient neural network architecture adapted to the specific characteristics of black hole imaging data to enhance regression performance.

To address these challenges, we propose a Multiscale Adaptive Network (MANet) that enhances modeling capacity and prediction robustness for black hole image parameter regression. MANet adopts a ResNet backbone and integrates two key modules in intermediate layers\cite{ResNet}: a Multiscale Enhancement Feature Pyramid (MEFP) and an Adaptive Channel Attention (ACA) mechanism. MEFP fuses spatial features across multiple scales to better represent multilayered structures such as photon rings and accretion disks, mitigating information loss caused by downsampling. ACA dynamically adjusts channel weights to enhance feature responses in important physical regions, filtering redundant information and improving feature discrimination. Ablation studies indicate that both modules contribute to performance improvements independently, and their combination yields further gains. Experiments under simulated observational noise show that MANet maintains stable performance and generalization. Even with limited GRRT training data, MANet effectively captures complex physical features, demonstrating its applicability in data-constrained scenarios.

This paper is organized as follows. Sec. 2 introduces the proposed MANet, detailing its overall architecture and two key components: the ACA mechanism and the MEFP. MANet aims to overcome the challenges to high-precision physical parameter regression posed by limited observational resolution, costly observations leading to sparse and uneven data, and the intrinsic complexity of black hole images. Sec. 3 presents the experimental settings and comprehensive analyses. It includes the dataset and preprocessing pipeline, evaluation metrics, and a comparison of different learning rate scheduling strategies. Visual interpretations of attention-induced activations are provided to examine how the model focuses on key relativistic structures such as photon rings and accretion disks. In addition, ablation studies are conducted to quantify the contributions of individual modules, while robustness evaluations under simulated observational noise further assess the model’s stability in astrophysically realistic scenarios. Sec. 4 concludes the paper and outlines future directions.

\section{Proposed method}

\begin{figure*}[htbp]
    \centering
    \includegraphics[width=\linewidth]{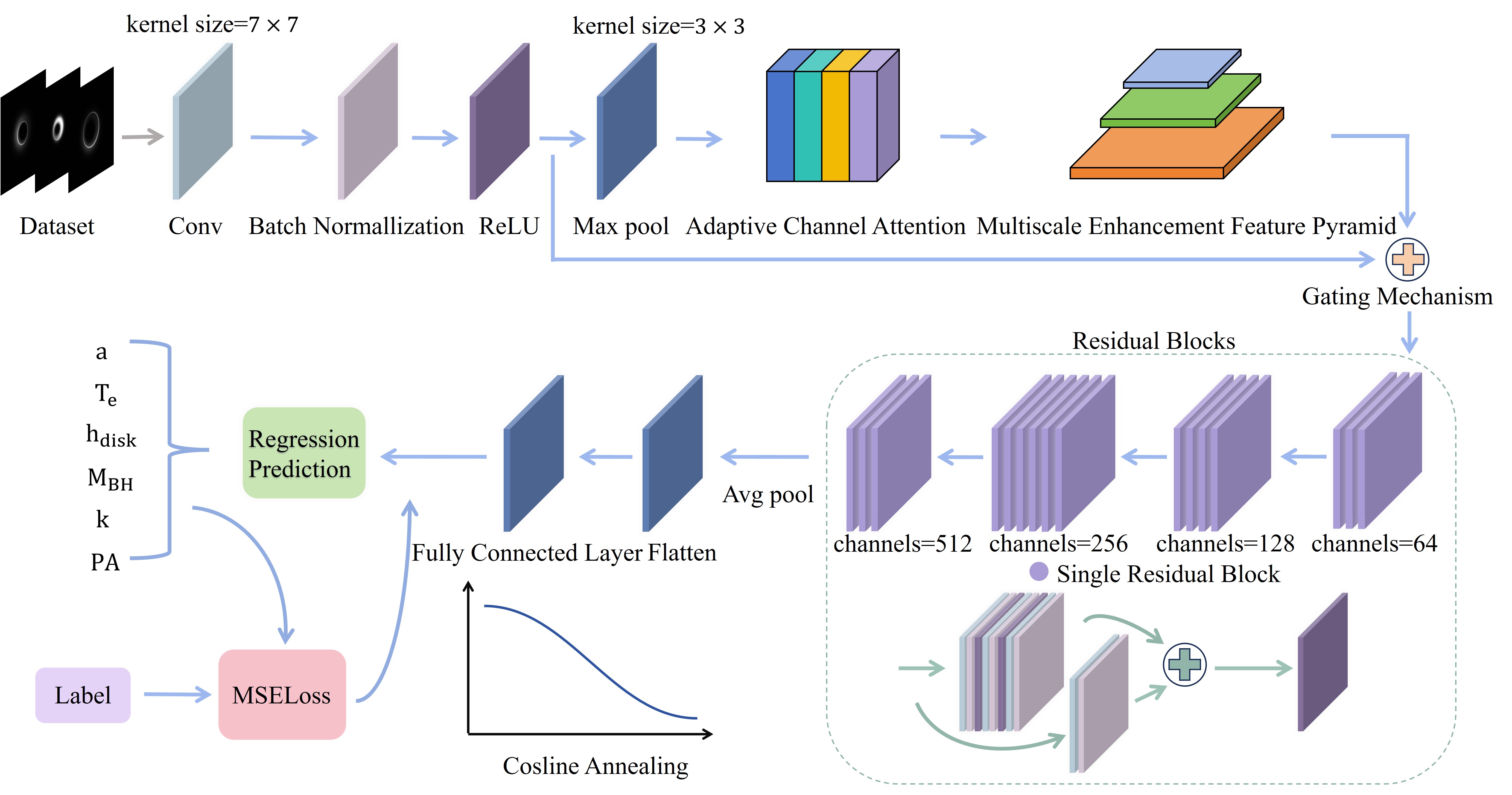}
    \caption{Structure of the proposed MANet for physical parameter regression from black hole images.}
\end{figure*}

To accurately invert physical parameters from black hole images, we propose the MANet, a deep convolutional architecture built upon a ResNet backbone. As illustrated in Figure 1, MANet retains the hierarchical feature extraction capabilities of ResNet, while embedding two specialized modules: ACA and MEFP. The ACA module adaptively highlights informative feature channels, while MEFP captures multiscale spatial structures crucial for robust physical parameter regression. The enhanced features are fused with the original residual representation via a learnable channel-wise gating mechanism, enabling the model to adaptively balance raw and enriched information for robust parameter regression.

\subsection{Adaptive Channel Attention}

In the context of black hole image parameter regression, the complexity and variability of physical features—such as relativistic beaming-induced brightness asymmetries and jet boundaries—require the model to selectively enhance channels that encode critical diagnostic information. This motivates the introduction of an adaptive channel attention module that adaptively learns to assign dynamic importance weights to feature channels based on global contextual information extracted via global average pooling. Figure 2 illustrates the internal structure of the ACA module.

\begin{figure*}[htbp]
    \centering
    \includegraphics[width=\linewidth]{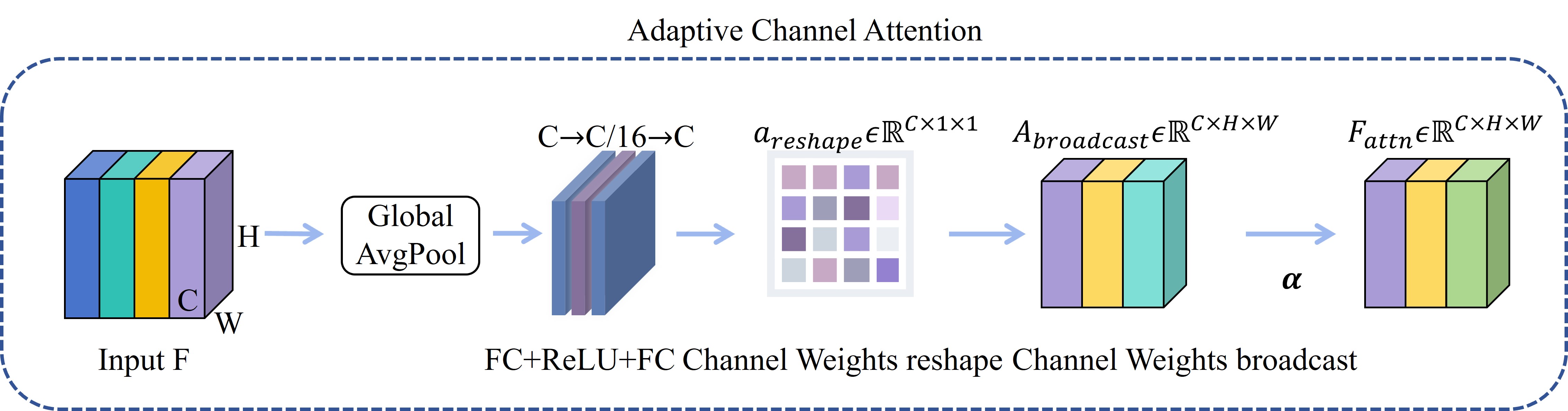}
    \caption{Architecture of the ACA module.}
\end{figure*}

Given an input feature map $F \in \mathbb{R}^{C \times H \times W}$, we compute a channel-wise descriptor by averaging over spatial dimensions:

\begin{equation}
    z = \text{GlobalAvgPool}(F) \in \mathbb{R}^{C}
\end{equation}

This descriptor is passed through a fully connected layer to produce channel-wise attention weights:

\begin{equation}
    a = \text{FC}(z) \in \mathbb{R}^{C}
\end{equation}

After reshaping to $\mathbb{R}^{C \times 1 \times 1}$, the attention weights are broadcast across spatial dimensions and applied to the input feature map:

\begin{equation}
    F_{\text{attn}} = F \cdot (a \cdot \alpha)
\end{equation}
where $\alpha \in [0, 1]$ is a learnable scaling factor controlling the attention strength. This adaptive reweighting enables the model to dynamically emphasize channels corresponding to physically meaningful regions, improving interpretability and robustness under complex astrophysical conditions.

The attention-enhanced feature map $F_{\text{attn}}$ is then passed through the MEFP module to extract rich spatial information across multiple scales, producing $F_{\text{MEFP}}$.

To balance the original features and the enhanced multiscale features, a learnable gating parameter $g \in \mathbb{R}^C$ is introduced. The final output $F_{\text{gated}}$ is computed as:

\begin{equation}
    F_{\text{gated}} = \sigma(g) \cdot F_{\text{MEFP}} + (1 - \sigma(g)) \cdot F
\end{equation}
where $\sigma(\cdot)$ denotes the sigmoid activation applied channel-wise. This gating mechanism allows the network to adaptively fuse the original and enhanced features, ensuring that the most informative representation is emphasized for accurate black hole parameter regression.

\subsection{Multiscale Enhancement Feature Pyramid}

Black hole images generated by GRRT simulations contain structural features spanning multiple spatial scales. The photon ring appears as a thin, high-contrast boundary, whereas the accretion disk manifests broader, smoother intensity variations. Additionally, relativistic beaming and gravitational lensing create complex emission patterns distributed across different receptive fields. When extracting features of certain key physical parameters that are not clearly observable—such as black hole spin and disk inclination—it is crucial to capture information at both fine and coarse spatial scales.

To address this challenge, we propose the MEFP module. Given an input feature map $F_{attn} \in \mathbb{R}^{C\times H\times W}$ produced by the preceding attention mechanism, where $C$ is the number of channels and $H, W$ denote spatial dimensions, MEFP employs three parallel convolutional branches with distinct kernel sizes to capture multiscale information.The multi-scale feature extraction design of MEFP is shown in Figure 3.

\begin{figure*}[htbp]
    \centering
    \includegraphics[width=\linewidth]{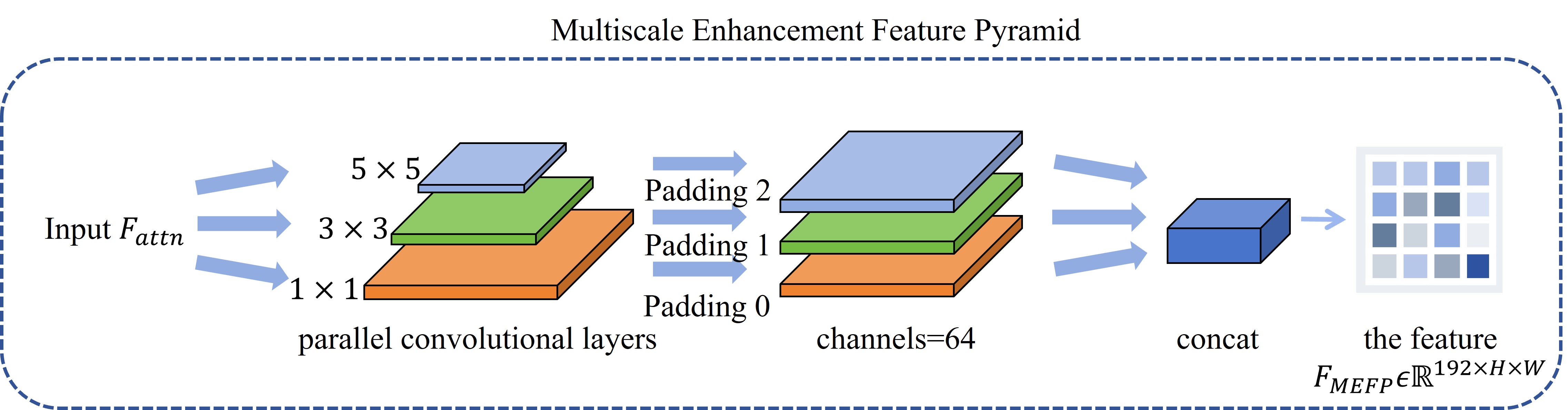}
    \caption{Architecture of the MEFP module.}
\end{figure*}

The first branch applies a $1 \times 1$ convolution with weights $W_1$ and bias $b_1$ , followed by a nonlinear activation $\sigma(\cdot)$ (e.g., ReLU), producing

\begin{equation}
    F_1=\sigma(W_1 * F_{attn} + b_1), F_1 \in \mathbb{R}^{C' \times H \times W}
\end{equation}

This branch preserves local spatial details critical for detecting sharp edges such as the photon ring.

The second branch captures medium-scale context by sequentially applying a $1 \times 1$ convolution $W_2^{(1)}, b_2^{(1)}$ for channel reduction, followed by a $3 \times 3$ convolution $W_2^{(3)}, b_2^{(3)}$ with padding 1 to maintain spatial dimensions:

\begin{equation}
    F_2 = \sigma(W_2^{(3)} * \sigma(W_2^{(1)} * F_{attn} + b_2^{(1)}) + b_2^{(3)}), F_2 \in \mathbb{R}^{C' \times H \times W}
\end{equation}

This branch extracts features related to mid-scale spatial patterns, such as the photon ring’s thickness and local asymmetries.

The third branch extracts coarse-scale features by applying a $1 \times 1$ convolution $W_3^{(1)}, b_3^{(1)}$, followed by a $5\times 5$ convolution $W_3^{(5)}, b_3^{(5)}$ with padding 2:

\begin{equation}
    F_3 = \sigma(W_3^{(5)} * \sigma(W_3^{(1)} * F_{attn} + b_3^{(1)}) + b_3^{(5)}), F_3 \in \mathbb{R}^{C' \times H \times W}
\end{equation}

This branch captures large receptive field information encompassing broad emission patterns and global spatial context of the accretion disk.

Because the convolutional branches use kernel-size-specific zero-padding (0 for $1 \times 1$\, 1 for $3\times 3$\, and 2 for $5\times 5$), all output feature maps $F_1, F_2, F_3$ retain identical spatial dimensions $H\times W$. This alignment enables their direct concatenation along the channel dimension:

\begin{equation}
    F_{concat} = Concat[F_1, F_2, F_3] \in \mathbb{R}^{3C' \times H \times W}
\end{equation}

Finally, a $1 \times 1$ convolution with weights $W_f$ and bias $b_f$ fuses the concatenated features and reduces the channel dimension back to $C$:

\begin{equation}
    F_{MEFP} = \sigma(W_f * F_{concat} + b_f), F_{MEFP} \in \mathbb{R}^{C\times H\times W}
\end{equation}

By integrating information across multiple receptive fields, MEFP mitigates the loss of critical spatial details caused by fixed kernel sizes and downsampling. It preserves fine-grained structures like the sharp photon ring edges while simultaneously capturing broader accretion disk morphology. This enriched feature representation enhances the model’s ability to regress black hole physical parameters accurately, reflecting the complex spacetime geometry and accretion physics encoded in the images.

\section{Experiments and analysis}

\subsection{Dataset and Pre-processing}

To construct a dataset tailored for black hole image parameter regression, we employ GRRT simulations to generate high-fidelity synthetic images that faithfully capture radiation from accretion flows under strong gravitational fields. GRRT enables precise treatment of light propagation effects such as gravitational lensing and Doppler beaming, ensuring that the simulated images are consistent with realistic observational conditions, such as those of the M87 black hole.

Each image in our dataset simulates a 230 GHz observation with a field of view of $128 \times 128$ microarcseconds and a $256 \times 256$ pixel resolution, reflecting commonly adopted settings in GRRT simulations related to the Event Horizon Telescope. The observer inclination is fixed at $163^\circ$, and the total flux of each image is normalized to 0.5 Jy for consistency.

\begin{table*}[htbp]
\caption{Physical Parameters Sampled for GRRT-Based Black Hole Image Simulation and Their Value Ranges}
\centering
\begin{tabular}{ccc}
\toprule
Parameter & Symbol & Range \\ 
\midrule
Black hole spin & $a$ & $[-1,1]$ \\
Black hole mass & $M_{BH}$ & $[2\times 10^9M_{\odot},10^{10}M_{\odot}]$ \\
Electron temperature & $T_e$ & $[10^9K,10^{12}K]$ \\
Accretion disk thickness & $h_{disk}$ & $[0.1,0.8]$ \\
Keplerian factor & $k$ & $[0,1]$ \\
Position angle & $PA$ & $[0^{\circ},360^{\circ}]$ \\
\bottomrule
\end{tabular}
\end{table*}

To ensure broad coverage of the physical parameter space, we uniformly sample six key parameters: black hole spin, mass, electron temperature, disk thickness, Keplerian velocity factor, and position angle, while randomly setting the disk rotation direction. In total, we generate 2157 images, each annotated with corresponding ground-truth physical parameters, forming a comprehensive dataset suitable for training and evaluating regression models. The ranges of these physical parameters are listed in Table 1, which provides an overview of the sampling space for GRRT-based simulation.

To ensure effective learning in black hole image parameter regression, it is critical to address the disparity in scale and units across physical parameters. Directly feeding raw values into the model may lead to instability or gradient inefficiency due to magnitude imbalances. To mitigate this, we adopt a standardization strategy that normalizes each parameter dimension to a standard normal distribution with zero mean and unit variance. This follows established practices in scientific data regression and ensures scale consistency across parameters.

Given the $k$-th physical parameter set ${x_1,x_2,\dots,x_n}$, we compute the empirical mean and standard deviation as:

\begin{equation}
    \mu_k = \frac{1}{n} \sum_{i=1}^n x_i,\quad \sigma_k = \sqrt{\frac{1}{n-1} \sum_{i=1}^n (x_i - \mu_k)^2}
\end{equation}

Each sample is then standardized by:

\begin{equation}
    Z_r = \frac{x_r - \mu_r}{\sigma_r},\quad r \in \{1,2,\dots,6\}
\end{equation}

This transformation not only improves training stability and convergence by eliminating scale-induced biases but also enhances the model’s ability to learn joint representations across heterogeneous physical quantities. Importantly, the transformation is reversible—original physical values can be recovered from predicted outputs using stored normalization statistics, preserving interpretability and physical significance.

In our implementation, normalization is applied independently to each parameter category to avoid dominance by parameters with larger magnitudes. This parameter-wise standardization ensures balanced gradient flow and improves the model’s capacity to extract cross-scale physical features relevant to GRRT-simulated black hole images.

To facilitate model training and evaluation, the dataset is partitioned into training, validation, and test subsets. The dataset comprises 2157 GRRT-simulated black hole images annotated with corresponding physical parameters.

Following standard machine learning protocols, we allocate 1725 samples (80.0\%) for training, 216 samples (10.0\%) for validation, and 216 samples (10.0\%) for testing. The training set is used for iterative model optimization, the validation set guides hyperparameter tuning, and the test set provides an independent assessment of the model’s generalization performance.

To prevent distributional shifts and ensure statistical consistency across subsets, we adopt stratified random sampling. This ensures that critical physical parameters—such as black hole mass and accretion disk thickness—are uniformly distributed across the splits, thereby minimizing potential evaluation bias and preserving the representativeness of each subset. This ensures fair evaluation of model generalization and supports reliable performance benchmarking in subsequent experiments.

\subsection{Evaluation indicators}

In black hole image parameter regression tasks, accurately assessing model performance and interpretability requires appropriate evaluation metrics. The coefficient of determination, denoted as $R^2$, is widely adopted as the primary metric because it effectively quantifies the proportion of variance in the target parameters explained by the model\cite{R2-1,R2-2}.

Mathematically, $R^2$ is defined as:

\begin{equation}
    R^2 = 1 - \frac{\sum_{i=1}^n (y_i - \hat{y}_i)^2}{\sum_{i=1}^n (y_i - \bar{y})^2}
\end{equation}
where $y_i$ is the true value of the $i$-th sample, $\hat{y}_i$ is the predicted value from the model, and $\bar{y} = \frac{1}{n} \sum_{i=1}^n y_i$ is the mean of the true values over all samples.

This metric provides an interpretable measure of goodness-of-fit, with values closer to 1 indicating better explanatory power of the model on the variability of physical parameters.

\subsection{Learning Rate Scheduling Strategy Selection: Cosine Annealing}

In GRRT-simulated black hole image regression tasks, we compared four prevalent learning-rate schedules—constant rate, step decay, exponential decay and cosine annealing—and observed that cosine annealing substantially outperformed the others in validation $R^2$. The initial gentle decay phase of cosine annealing preserves a high learning rate for extensive global exploration of the intricate, multiscale spacetime features in black hole images, thereby avoiding premature convergence to suboptimal solutions. Its mid-training rapid reduction concentrates parameter updates in promising regions, accelerating convergence and improving optimization efficiency. Finally, the late slow taper facilitates precise fine-tuning, enhancing the accuracy of subtle spacetime curvature estimations.

Under a unified experimental setup with an initial learning rate of $9 \times 10^{-4}$, $500$ training epochs and identical train/validation splits, Table 2 summarizes the $R^2$ performance of each learning rate schedule across six target parameters and Figure 4 illustrates the learning rate and loss dynamics for parameter $a$, shown here as a representative case.

\begin{table*}[h]
\caption{$R^2$ Scores for Various Learning Rate Schedules in Black Hole Parameter Regression}
\centering
\begin{tabular}{ccccccc}
\toprule
 & $a$ & $T_e$ & $h_{disk}$ & $M_{BH}$ & $k$ & $PA$ \\
\midrule
Constant Rate & $0.9866$ & $0.8253$ & $0.8910$ & $0.9980$ & $0.9691$ & $0.8987$ \\
Step Decay & $0.9881$ & $0.9836$ & $0.9653$ & $0.9987$ & $0.9583$ & $0.8565$ \\
Exponential Decay & $0.9582$ & $0.9943$ & $0.9841$ & $0.9986$ & $0.9534$ & $0.8182$ \\
Cosine Annealing(\textbf{ours}) & $\mathbf{0.9906}$ & $\mathbf{0.9969}$ & $\mathbf{0.9927}$ & $\mathbf{0.9993}$ & $\mathbf{0.9725}$ & $\mathbf{0.9602}$\\
\bottomrule
\end{tabular}
\end{table*}

As shown in Table 2, cosine annealing consistently achieves the highest $R^2$ scores across all six physical parameters. For $T_e$ and $h_{disk}$—parameters that are particularly sensitive to high-frequency image details—the relative improvements reach 1.3\% and 1.1\%, respectively, compared to the best alternative. Moreover, for $PA$, which encodes orientation-related information, cosine annealing surpasses step decay by more than 10\%. These improvements demonstrate that the cosine learning rate schedule is better suited to the intricate, multi-modal optimization landscape characteristic of GRRT-based regression tasks.

\begin{figure*}[htbp]
    \centering
    \includegraphics[width=\linewidth]{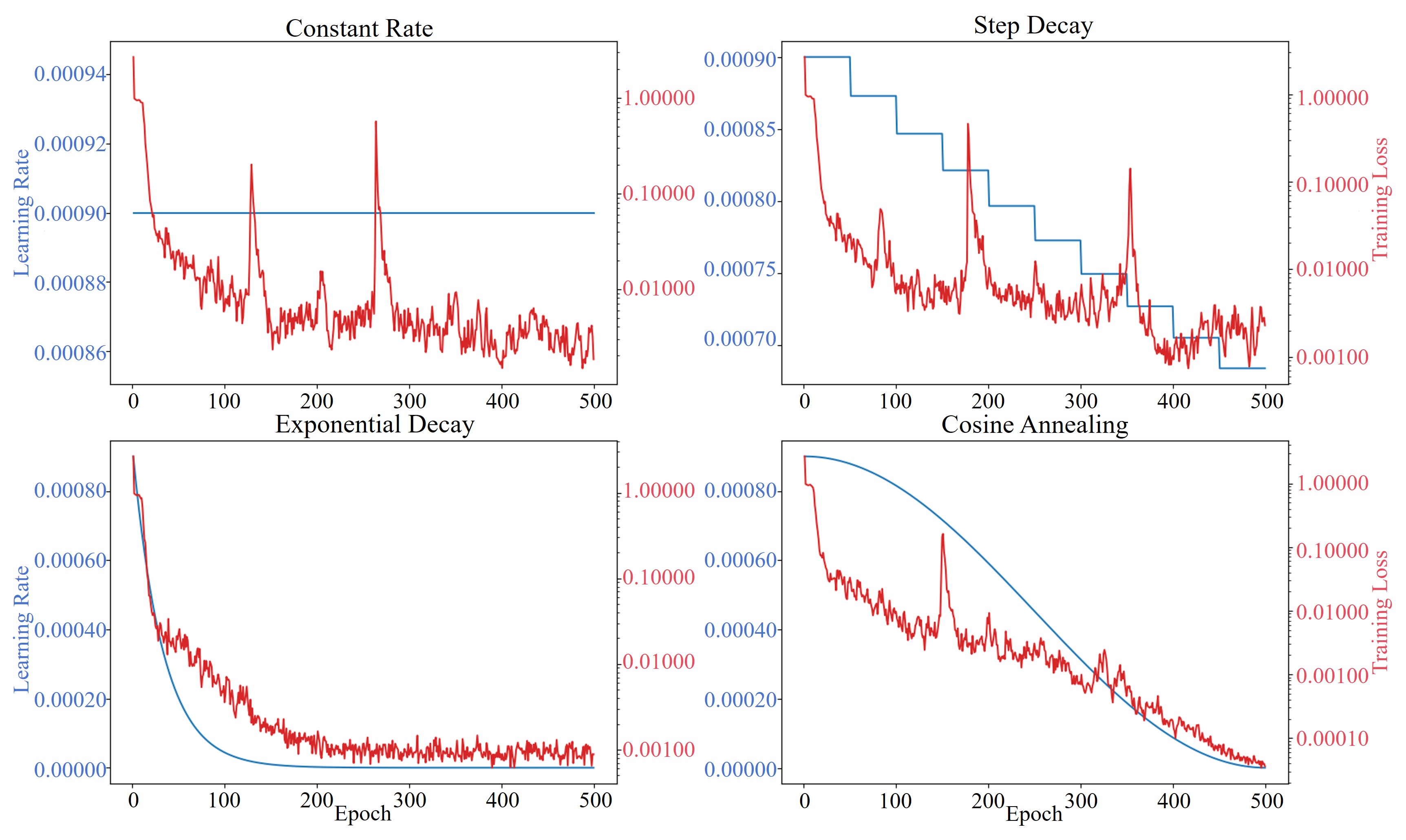}
    \caption{Training Dynamics of Learning Rate and Loss Under Different Scheduling Strategies.(Shown for Parameter $a$) Loss is plotted on a logarithmic scale to better illustrate convergence across several orders of magnitude.}
\end{figure*}

Figure 4 shows the training dynamics for parameter $a$ as a representative example. Among the four learning rate scheduling strategies compared, the final loss values are as follows: constant rate achieves a loss of $1.838 \times 10^{-3}$, step decay $2.285 \times 10^{-3}$, exponential decay $8.843 \times 10^{-4}$, while cosine annealing achieves the lowest loss of $3.618 \times 10^{-5}$. This indicates that cosine annealing enables more effective optimization and better convergence in black hole parameter regression tasks.

\subsection{Visual Interpretation of the Attention Mechanism}

To qualitatively assess how the adaptive channel attention mechanism influences spatial feature localization in black hole image regression, we employ Gradient-weighted Class Activation Mapping (Grad-CAM)\cite{Grad-CAM,CAM2,CAM3} to generate activation heatmaps. These visualizations highlight the input regions contributing most to the model's predictions, offering insight into how attention reshapes feature weighting in alignment with physical structures. This is particularly relevant for GRRT-simulated black hole images, which exhibit bright compact features such as the photon ring and inner accretion disk, surrounded by diffuse backgrounds.

\begin{figure*}[htbp]
    \centering
    \includegraphics[width=\linewidth]{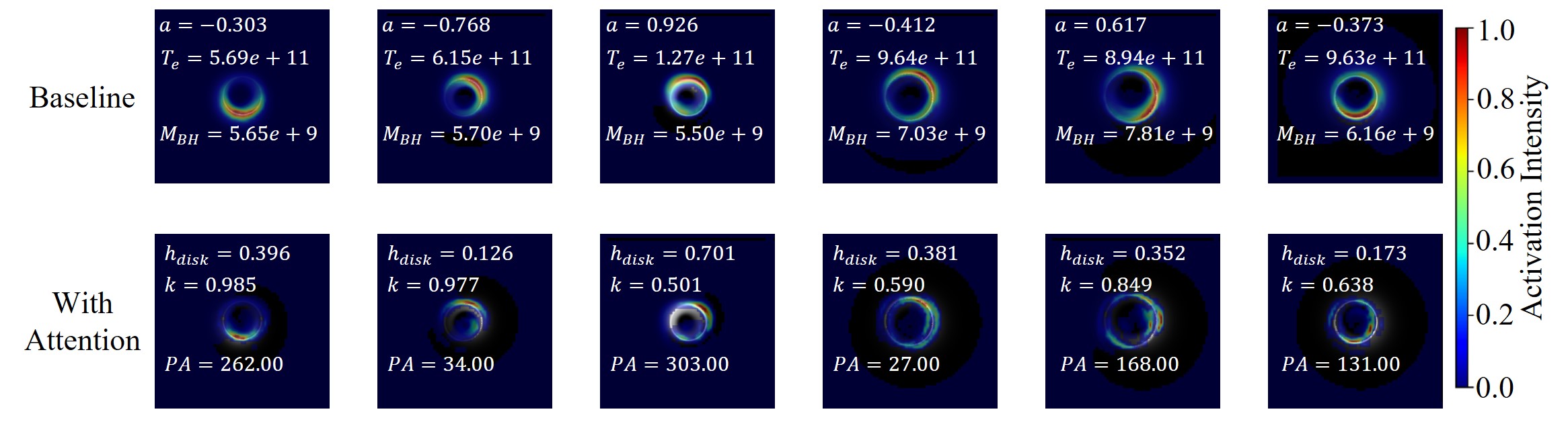}
    \caption{Activation Heatmaps Comparing Baseline and Attention-Enhanced Models for Black Hole Image Parameter Regression. Each column presents two heatmaps derived from the same input image and identical ground truth parameters. For readability, the upper image (baseline model) displays three of the parameters, while the lower image (with attention model) displays the remaining three. A color scale representing relative activation intensity is included to facilitate interpretation of the spatial focus of each model.}
\end{figure*}

As shown in Figure 5, comparing activation patterns between the baseline and attention-enhanced models reveals meaningful differences. The baseline model already demonstrates reasonable localization of the black hole structure, typically activating circular or wide ring-shaped regions. However, with the introduction of the attention mechanism, the model produces more structured and spatially selective heatmaps. The broad activation ring becomes thinner and more sharply delineated, occasionally fragmenting into arc-like segments that more precisely trace the morphology of the photon ring and the high-emission regions of the accretion disk.

These localized, high-gradient zones encode key physical information such as black hole spin, viewing angle, and accretion dynamics. By concentrating attention on these critical structures, the model enhances its ability to capture physically informative representations, leading to improved regression accuracy and stronger alignment between learned features and the underlying spacetime geometry.

\subsection{Ablation Experiments}

To evaluate the synergistic contribution of the Multiscale Enhancement Feature Pyramid MEFP and the attention module to the performance of black hole image parameter regression, we conducted a series of ablation experiments. Starting from the complete network, we successively removed the MEFP module and the attention mechanism to construct reduced variants. All experiments were carried out under the same training configuration and data partition as described in the previous sections, with the coefficient of determination $R^2$ used as the evaluation metric across six physical parameters derived from GRRT-simulated black hole images. The results of these experiments are summarized in Table 3, enabling a controlled assessment of how each architectural component influences the model’s overall regression performance.

\begin{table*}[htbp]
\caption{$R^2$ performance comparison of MANet and ablated variants on simulated black hole datasets}
\centering
\begin{tabular}{ccccccc}
\toprule
 & $a$ & $T_e$ & $h_{disk}$ & $M_{BH}$ & $k$ & $PA$ \\
\midrule
ResNet50 & $0.9359$ & $0.9284$ & $0.9708$ & $0.9986$ & $0.9556$ & $0.8569$ \\
ResNet50+MEFP & $0.9887$ & $0.9819$ & $0.9871$ & $0.9992$ & $0.9603$ & $0.9181$ \\
ResNet50+ACA & $0.9864$ & $0.9759$ & $0.9912$ & $0.9990$ & $0.9561$ & $0.9459$ \\
MANet(\textbf{ours}) & $\mathbf{0.9906}$ & $\mathbf{0.9969}$ & $\mathbf{0.9927}$ & $\mathbf{0.9993}$ & $\mathbf{0.9725}$ & $\mathbf{0.9602}$\\
\bottomrule
\end{tabular}
\end{table*}

As presented in Table 3, both the MEFP module and the channel attention mechanism independently contribute to performance gains across all six regression targets. When the MEFP is added to the baseline ResNet50, the $R^2$ score for electron temperature $T_e$ increases from 0.9284 to 0.9819—a significant gain of 0.0535. Similarly, $h_{disk}$ improves from 0.9708 to 0.9871, highlighting the module’s strength in capturing hierarchical spatial cues. This suggests that MEFP is particularly effective for parameters governed by multiscale morphological gradients, as commonly found in GRRT-generated accretion flows.

The attention module, on the other hand, delivers the most pronounced improvements in parameters like spin $a$ and position angle $PA$. For example, the $R^2$ for $PA$ rises from 0.8569 to 0.9459—an absolute increase of 0.089—indicating its ability to enhance directional signals and suppress irrelevant background emissions through adaptive channel-wise reweighting.

When both modules are integrated into the full MCNet architecture, the model achieves consistent and superior results across all parameters. Notably, the $R^2$ values reach 0.9906 for $a$, 0.9969 for $T_e$, and 0.9602 for $PA$, outperforming all ablated variants. This synergy between multiscale feature abstraction and attentive refinement enables the model to better disentangle complex emission patterns, leading to more robust and physically accurate parameter estimations under realistic image conditions.

\subsection{Robustness Evaluation Under Simulated Observational Noise}

In practical astronomical observations, imaging data inevitably suffer from various noise sources due to instrument limitations and environmental factors. To realistically simulate these unavoidable noise disturbances and further evaluate the robustness of the proposed framework under real-world imaging conditions, we conducted additional experiments by introducing Gaussian noise to the test images. Gaussian noise is a common artifact in image acquisition, particularly in high-sensitivity astronomical observations, where signal instability and instrumental interference are prevalent. In this study, we injected zero-mean Gaussian noise with standard deviations of 10 and 20—referred to as Gauss10 and Gauss20, respectively—into the test set to simulate varying levels of signal degradation typically encountered in practical scenarios\cite{gauss1,gauss2,gauss3,gauss4}.

Figure 6 presents representative examples before and after noise injection, while Table 4 reports the $R^2$ scores for different models across six physical parameters under each noise condition.

\begin{figure*}[htbp]
    \centering
    \includegraphics[width=0.6\linewidth]{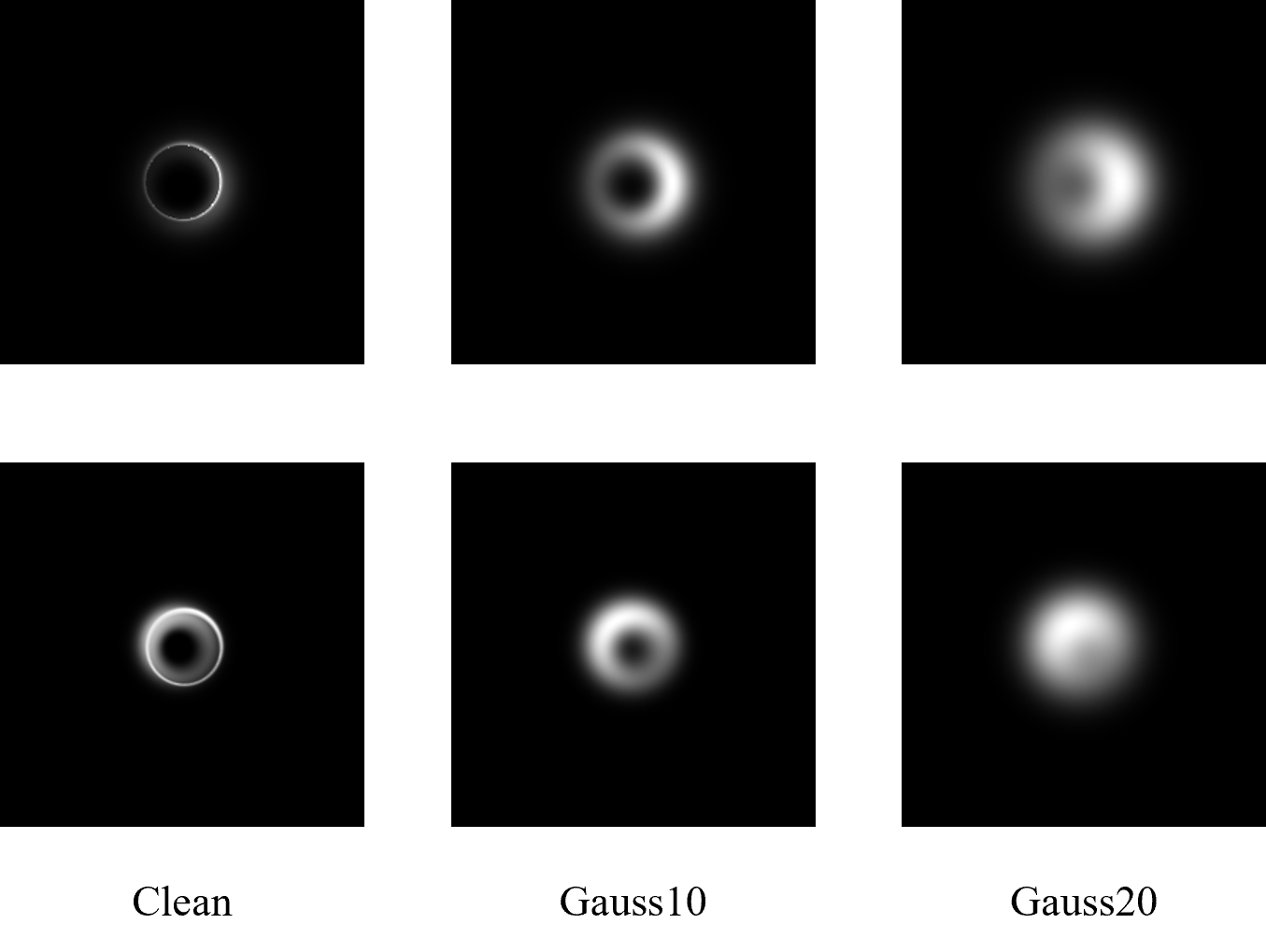}
    \caption{Representative Examples of Black Hole Images Before and After Gaussian Noise Injection}
\end{figure*}

\begin{table*}[htbp]
\caption{Comparison of Model Parameter Estimation $R^2$ Scores Under Simulated Observational Noise}
\centering
\begin{tabular}{ccccccc}
\toprule
 & $a$ & $T_e$ & $h_{disk}$ & $M_{BH}$ & $k$ & $PA$ \\
\midrule
ResNet50$(10\mu a s)$ & $0.9444$ & $0.8876$ & $0.9538$ & $0.9927$ & $0.9754$ & $0.8376$ \\
\textbf{ours}$(10\mu a s)$ & $\mathbf{0.9641}$ & $\mathbf{0.9976}$ & $\mathbf{0.9775}$ & $\mathbf{0.9960}$ & $\mathbf{0.9807}$ & $\mathbf{0.8998}$\\
ResNet50$(20\mu a s)$ & $0.6868$ & $0.8862$ & $-0.0949$ & $0.9742$ & $0.9491$ & $0.8507$ \\
\textbf{ours}$(20\mu a s)$ & $\mathbf{0.7029}$ & $\mathbf{0.9862}$ & $\mathbf{-0.0060}$ & $\mathbf{0.9744}$ & $\mathbf{0.9580}$ & $\mathbf{0.8601}$\\
\bottomrule
\end{tabular}
\end{table*}

Experimental results, as shown in Table 4, indicate that under moderate noise conditions (Gauss10), the proposed MANet maintains high regression accuracy across all parameters, demonstrating robustness against mild perturbations. As the noise level increases to Gauss20, MANet continues to outperform the baseline model in most cases, demonstrating robust generalization. However, the regression accuracy of the disk height $h_{disk}$ declines significantly, with the $R^2$ score dropping to $-0.0060$, suggesting that this physical parameter is particularly sensitive to noise interference. This sensitivity is mainly due to the inherently subtle and fine-grained morphological features associated with $h_{disk}$ in the GRRT images, which become increasingly obscured by higher noise levels, as visually illustrated in Figure 6. In contrast, parameters such as the electron temperature $T_e$ and black hole mass $M_{BH}$ exhibit relative robustness, maintaining high $R^2$ scores even under severe noise conditions. These observations highlight the intrinsic difficulty of accurately estimating parameters characterized by delicate image features in noisy observational environments.

\section{Conclusion}

In this study, we propose MANet, a ResNet-based regression network that integrates a MEFP and an ACA module to improve the regression of black hole image parameters. Traditional approaches, such as forward modeling and Bayesian inference, typically rely on exhaustive sampling of the parameter space or computationally intensive probabilistic computations, which can be inefficient and poorly scalable in high-dimensional or data-scarce settings. By contrast, the MEFP module captures fine-grained physical structures through multiscale feature fusion, while the ACA module highlights critical regions such as the photon ring and brightness asymmetries using a dynamic attention mechanism. Leveraging deep learning, MANet learns complex nonlinear mappings directly from data, improving estimation accuracy while significantly reducing inference time. Experimental results show that both MEFP and ACA independently contribute to performance gains, and their combination further enhances accuracy and robustness. Even under conditions of limited data and simulation noise, MANet exhibits strong generalization ability and stability, demonstrating the effectiveness of combining multiscale feature fusion with adaptive attention for high-precision astrophysical regression tasks. In the current work, we focus on the regression of six representative parameters. In future research, we plan to incorporate temporal features and uncertainty quantification into the network architecture to improve reliability and interpretability, further enhancing the model’s applicability to real observational scenarios.

\section*{Data availability statement}
The data that support the findings of this study are openly available in Science Data Bank at 

\href{https://www.doi.org/10.5281/zenodo.15846647}{https://www.doi.org/10.5281/zenodo.15846647}

\nolinenumbers

\clearpage

\addcontentsline{toc}{chapter}{Appendix A: Additional Grad-CAM Visualizations}
\section*{Appendix A: Additional Grad-CAM Visualizations}

In addition to the six examples presented in the main text, this appendix provides supplementary Grad-CAM visualizations on several additional samples. These expanded examples offer a more comprehensive understanding of the attention mechanism’s behavior, including cases with clear focus as well as ambiguous or challenging scenarios. Figure 7 illustrates these additional Grad-CAM maps, which further demonstrate the interpretability and robustness of the model’s attention mechanism by highlighting both typical and difficult cases.

\begin{figure*}[htbp]
    \centering
    \includegraphics[width=\linewidth]{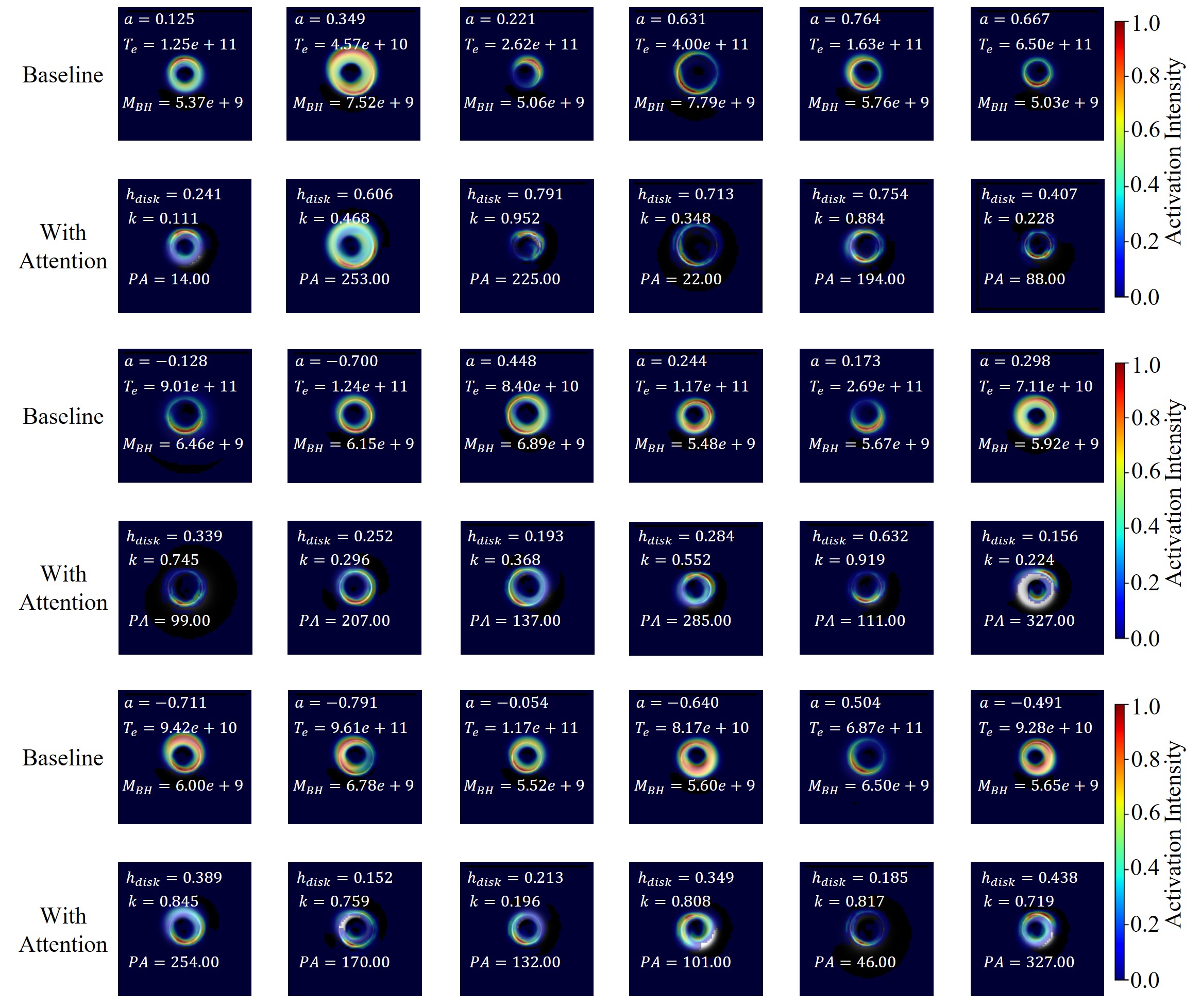}
    \caption{Additional Grad-CAM visualizations illustrating the attention maps on diverse samples, including typical cases and challenging scenarios to further demonstrate the interpretability and robustness of the model.}
\end{figure*}

\end{document}